\documentclass[showpacs,twocolumn,pre,floatfix]{revtex4}
\usepackage{psfrag,epsfig,amsfonts,amssymb,amsmath}
\usepackage{dcolumn}

\usepackage[normalem]{ulem}
\usepackage{color}

\begin{document}
\newcommand{\RR}{{\mathbb R}}
\newcommand{\ord}{{\cal O}}
\newcommand{\LL}{{\mathrm{L}}}

\title{Chiral particle separation by a non-chiral micro-lattice}

\author{Lukas Bogunovic$^{1}$}
\author{Marc Fliedner$^{1}$}
\author{Ralf Eichhorn$^2$}
\author{Sonja Wegener$^{1}$}
\author{Jan Regtmeier$^{1}$}
\author{Dario Anselmetti$^{1}$}
\author{Peter Reimann$^{1}$}
\affiliation{$^{1}$Bielefeld University, Faculty of Physics, 33615 Bielefeld, Germany\\
$^2$Nordita, Royal Institute of Technology and Stockholm University, 
Roslagstullsbacken 23, SE-106 91 Stockholm, Sweden}

\begin{abstract}
We conceived a model experiment for a continuous separation 
strategy of chiral molecules (enantiomers) without the need 
of any 
chiral selector structure
or derivatization agents:
Micro-particles that only differ by their chirality are 
shown to migrate along different directions when driven 
by a steady fluid flow
through a square lattice of cylindrical posts.
In accordance with our numerical predictions,
the transport directions of the enantiomers depend
very sensitively on the orientation of the lattice 
relatively to the fluid flow.
\end{abstract}

\maketitle

Many building blocks of life 
(e.g. a ligand) 
are organic molecules 
of one specific chirality which interact very differently with other 
molecule species (e.g. a receptor)
due to
their handedness (enantiomers):
While one species 
may act as a drug, its chiral partner may be inert or 
even toxic \cite{chiral-grundlagen}.
Methods for separating enantiomers are thus of great importance in fields 
ranging from molecular biology and organic chemistry to pharmacy and agriculture.
Up to date, all those methods require suitable chiral derivatization agents 
or some other kind of chiral selector material or structure \cite{klassische-trennung}.
Their main disadvantage is that they have to be 
developed and precisely predesigned 
anew for every given chiral analyte.

\begin{figure}[h]
\epsfxsize=\columnwidth
\epsfbox{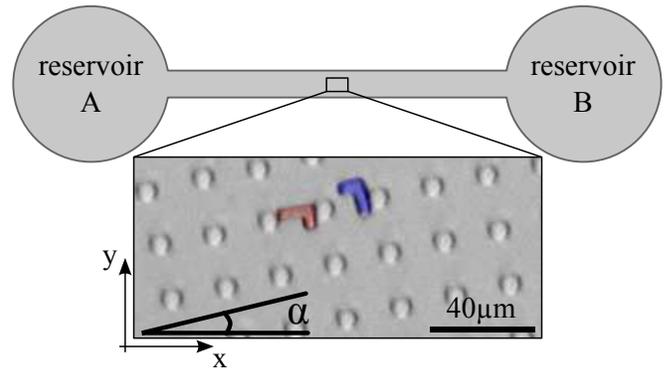}
\caption{Schematic top-view ($x$-$y$-plane)
of the microfluidic experiment 
with fluid- and particle-reservoirs at its ends.
The magnified inset
shows an optical micrograph 
image of the central, microstructured area
(square lattice of cylindrical post)
and two particles of opposite chirality.
A steady fluid flow in $x$-direction
is generated by a pneumatic pump (not shown),
maintaining a constant pressure 
difference between the reservoirs.
The angle $\alpha$ 
denotes the inclination
of the square lattice relatively to the 
fluid flow.
}
\label{fig1}
\end{figure}

Here, we put forward an alternative approach which does 
not require any such selector with a built in 
{\em structural} chirality, but instead exploits 
a {\em dynamical} chiral symmetry breaking.
The proposed model experiment from 
Fig. \ref{fig1} builds on previous separation 
strategies for {\em non-chiral} particles,
as conceived theoretically e.g. in Refs. \cite{theo1,theo2} 
and realized experimentally e.g. by
Austin, Sturm, and coworkers \cite{exp}.
Conceptually, our approach may be traced back to de Gennes, 
pointing out that small (but still macroscopic) objects of opposite chirality
generically must exhibit different transport directions when sliding 
(or rolling) down a solid plane or floating on a liquid surface under 
the action of a constant force \cite{degennes}.
More specifically, chiral particles which are propelled in a 
shear flow have been explored in several theoretical \cite{kim}
and experimental \cite{howard,chen,makino08,marcos} studies,
demonstrating that the right- and left-handed particles
experience a lift force of opposite sign in the vorticity 
direction of the flow.
For particles of one specific chirality,
the resulting deflection from the main flow 
direction and its dependence on the shear rate
was reported by Markino et al. for millimeter-sized 
chiral objects \cite{makino08},
and by Marcos et al. for micrometer-sized
helical bacteria \cite{marcos}.
A genuine separation of right- and left-handed, centimeter-sized 
crystals was demonstrated by Howard \cite{howard}. 
Another main theoretical concept established by Kostur et al.
exploits the interplay of thermal noise and of a flow field with spatially 
variable vorticity to separate stylized chiral ``molecules''
\cite{kostur}.
In a next step, Eichhorn theoretically predicted that
right- and left-handed particles are pumped with different velocities
through a microfluidic channel with asymmetric walls \cite{eichhorn}.

With our present work we further pursue yet 
another theoretical proposal \cite{speer},
namely to employ periodic potentials for separating 
chiral particles.
Our experimental realization of this concept 
is schematically illustrated in Fig.\ \ref{fig1},
featuring a central microstructured square
lattice array
extending over 1 cm in 
$x$-direction and 1 mm in $y$-direction,
which consists of cylindrical posts 
with $3.3\pm 0.2 \:\mu$m radius, 
$20.5\pm 0.2 \:\mu$m lattice constant,
and $6\pm 0.2\:\mu$m height ($z$-direction).
The entire device is filled with distilled water,
containing $\LL$- and $\Gamma$-shaped ``enantiomers''
in sufficiently high dilution so that their 
interaction can be safely neglected.
Their long and short axes are
$14.7\pm 0.3 \:\mu$m and 
$10.2\pm 0.3 \:\mu$m respectively, 
and their height ($z$-direction) 
is 3.5$\pm 0.3\: \mu$m.
Hence, they are small enough to edge their 
way through the lattice and large enough to 
prevent their swapping of chirality
by rotation about their long axis.
For a more detailed description of the
particle and microstructure fabrication
we refer to \cite{details}.
The experiment is actuated by a 
pneumatic
pump, maintaining a constant pressure 
difference of $p \simeq 15 \: \mathrm {mbar}$
between the reservoirs (see Fig. \ref{fig1}), 
resulting in a steady fluid flow 
of the order of 100$\,\mu$m/s.
During the experiment, a small part
of the structured area is observed 
with an inverted microscope (Zeiss Axiovert 200) 
and the particles are tracked by digital video 
microscopy.

As emphasized in \cite{speer}, there does not exist any
symmetry argument of why the $\LL$-and 
$\Gamma$-shaped enantiomers should travel with
identical mean velocities $\vec v$ through the 
square lattice when driven by a steady flow.
Following de Gennes \cite{degennes}, we thus can 
conclude that the two velocities will indeed be 
different 
unless we can name any 
{\em a priori} reasons to the contrary.
One important such reason identified in 
\cite{speer} are locking phenomena 
in the absence of thermal fluctuations 
\cite{theo2}.
Since thermal noise effects turn out to
be rather weak in our present experiment,
the velocities of the two enantiomers 
are expected to be strictly speaking 
different but still practically 
indistinguishable under many conditions.
To avoid such unwanted cases and rather 
optimize the experimental set-up {\em a priori},
we performed extensive numerical 
simulations \cite{f1}.
In particular, we numerically explored
many different (but still experimentally 
feasible) post diameters, lattice constants, 
particle dimensions, pressure differences, 
and inclination angels $\alpha$.
In the following, we immediately focus
on the resulting ``best case prediction'', 
from which we derived the set-up 
specified in Fig. \ref{fig1} and in the 
previous paragraph.

\begin{figure}[h]
\epsfxsize=\columnwidth
\epsfbox{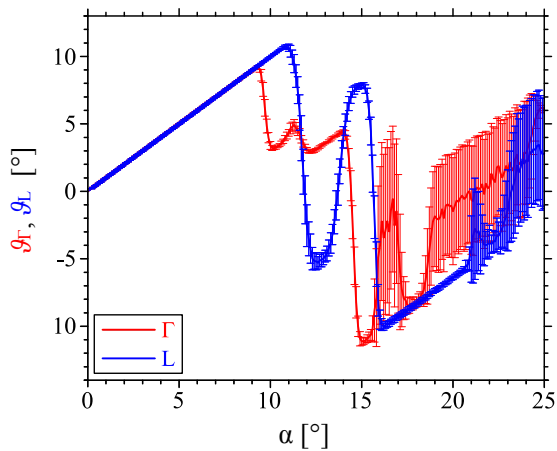}
\caption{Deflection angles $\vartheta_\Gamma$
and $\vartheta_\LL$ of the $\Gamma$- and 
$\LL$-shaped particles versus lattice 
inclination $\alpha$, 
predicted by means of numerical simulations
for parameters corresponding to the 
experimental set-up from Fig. \ref{fig1}.
The error bars arise by averaging over many 
initial conditions and realizations of the noise 
in combination with a long but finite temporal average.
}
\label{fig2}
\end{figure}

The quantity of foremost interest is
the time-averaged particle velocity 
$\vec v$,
and in particular the deflection angle
$\vartheta$ of this velocity from the
$x$-axis.
Those numerically predicted deflection
angles for both the $\LL$- and $\Gamma$-shaped 
particles versus the inclination angle $\alpha$ 
of the lattice are presented in Fig. \ref{fig2}.
For $\alpha< 9^\circ$, one of the above 
mentioned locking effects occurs: 
Both particle species essentially follow the
(1,0) lattice direction 
($\vartheta_\LL=\vartheta_\Gamma=\alpha$), 
apart from extremely small deviations 
(not resolvable in Fig. \ref{fig2}) due to thermal noise.
In contrast, for $\alpha>9^\circ$ quite appreciable
differences $\vartheta_{\LL}-\vartheta_{\Gamma}$ between
the two transport directions arise.
Beyond $\alpha=16^\circ$, the error bars 
of $\vartheta_\Gamma$ and beyond $\alpha=21^\circ$ 
also those of of $\vartheta_\LL$ dramatically increase
in our simulations.
The reason is that, depending on the initial conditions,
numerical solutions with different time-averaged velocities 
$\vec v$ may coexist in the absence of thermal noise.
Since thermal noise is rather weak in our system, 
the velocities still exhibit a large variance within
our quite long but still finite simulation times.
To avoid such an undesirably slow convergence, 
we confine ourselves to $\alpha\leq 16^\circ$.
For symmetry reasons it is furthermore sufficient
to consider $\alpha\geq 0$.

According to Fig. \ref{fig2}, the largest splitting angles
$|\vartheta_\LL-\vartheta_{\Gamma}|$ between
the $\LL$- and $\Gamma$-shaped particles
are expected around $\alpha=15^\circ$.
A slightly weaker splitting, but remarkably enough
with an opposite sign of $\vartheta_\LL-\vartheta_{\Gamma}$,
is expected around $\alpha=12.5^\circ$.
Therefore, we will focus in the experiments 
on these two most promising values 
$\alpha=15^\circ$ and $\alpha=12.5^\circ$
predicted by the theory \cite{f2}.

As already mentioned, we performed similar 
simulations as in Fig. \ref{fig2} for many 
other lattice and particle parameters.
In order to maximise $|\vartheta_\LL-\vartheta_{\Gamma}|$,
the parameters from Fig. \ref{fig2} seem to be (practically) 
optimal, but significant $|\vartheta_\LL-\vartheta_{\Gamma}|$
could still be obtained for a wide range of
substantially different parameters.

\begin{figure}[h]
\epsfxsize=\columnwidth
\epsfbox{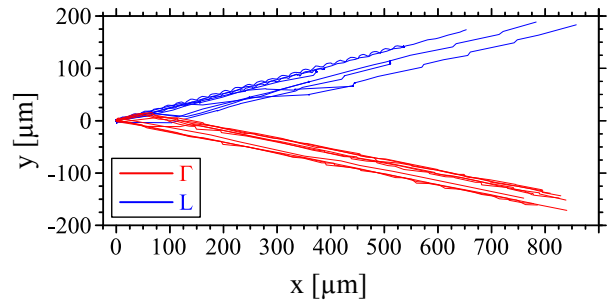}
\caption{Experimentally observed trajectories of 
$\LL$-
and $\Gamma$-shaped 
particles for the same set-up as in Fig. \ref{fig1} with 
{\boldmath$\alpha=15^\circ$}.
The total length of 
each trajectory depends on where the 
particle happened to enter and exit the visual 
field of the optical microcope.
}
\label{fig3}
\end{figure}

Fig. \ref{fig3} illustrates how the $\LL$- and $\Gamma$-shaped 
particles actually move in the experiment for 
$\alpha=15^\circ$.
The main point of Fig. \ref{fig3} is also the 
main point of our present paper: 
{\em In accordance with our numerical predictions, 
the two chiral particle species can indeed be separated 
from each other in our experimental microfluidic device 
with extremely high reliability.}
In fact, not a single particle with a ``wrong''
sign of $\vartheta$ was observed.

\begin{figure}[h]
\epsfxsize=\columnwidth
\epsfbox{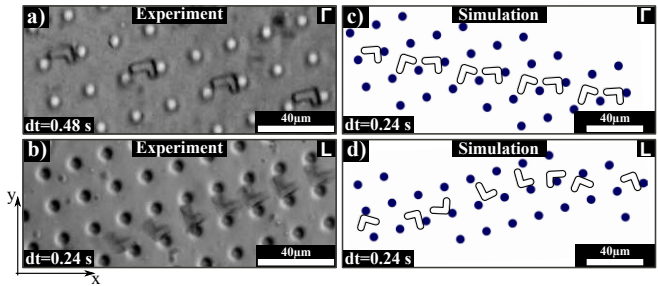}
\caption{
Left:
Stroboscopic images by superimposing 
several video microscopy snapshots
for one 
$\Gamma$-
or one $\LL$-shaped 
particle from Fig. \ref{fig3}
({\boldmath$\alpha=15^\circ$}).
The motion proceeds from left to right
at time steps $dt$.
Right:
Corresponding numerical simulations.
}
\label{fig4}
\end{figure}

\begin{figure}[h]
\epsfxsize=\columnwidth
\epsfbox{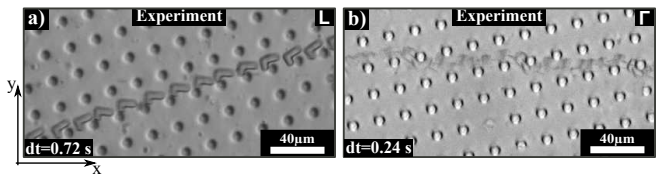}
\caption{(a):
Same as in Fig. \ref{fig4}b 
({\boldmath$\alpha=15^\circ$}),
but for a different video sequence of
an $\LL$-shaped particle and
a different time step $dt=$0.72\,s.
(b): Same as in Fig. \ref{fig6}a 
({\boldmath$\alpha=12.5^\circ$}),
but for a different video sequence of
a $\Gamma$-shaped particle
(the snapshots are blurred due to the
rapid particle motion).
}
\label{fig5}
\end{figure}

Fig. \ref{fig4} provides a more detailed picture of
how the particles travel through the square lattice.
The stroboscopic snapshots at time steps 
$dt=$0.48\,s in Fig. \ref{fig4}a demonstrate 
a very regular, time-periodic motion of the 
$\Gamma$-shaped particles in the experiment.
This behavior is extremely well reproduced 
by the simulations in Fig. \ref{fig4}c.
The reduced time step of $dt=$0.24\,s in 
Fig. \ref{fig4}c in addition reveals 
a quite pronounced rotational see-saw motion
(the corresponding experimental snapshots are
even more blurred than in Fig. \ref{fig5}b 
and are therefore not shown).
Due to thermal noise and imperfect 
spatial periodicity, also very rare
deviations from the behavior in Fig. 
\ref{fig4}a are observed (see also red 
trajectories in Fig. \ref{fig3}),
resulting in a mean experimental deflection
angle of $\vartheta_{\Gamma}= -10.7^\circ \pm 0.8^\circ$.

Turning to the $\LL$-shaped particles,
Fig. \ref{fig4}b evidences an extremely
``straight'' motion along one and 
the same inclined row of posts.
The simulation in Fig. \ref{fig4}d 
also shows short episodes of such a
straight motion, but additionally 
every now and then the particle 
bumps with its ``short leg'' against 
a post, turns around, 
collides with another post, 
and finally continues along
the next lower row of posts 
with another short episode 
of straight motion.
In fact, the complete experimental 
video, of which only a small part is covered by
Fig. \ref{fig4}b, also exhibits 
such transitions between rows of 
posts (see also the 
blue trajectories in Fig. \ref{fig3}),
but at a substantial smaller rate than in 
the simulations.
In addition, for some of the $\LL$-shaped particles 
we also observed a different type of motion in
the experiment, illustrated by Fig. \ref{fig5}a,
without a comparable theoretical counterpart.
(While the particles in Figs.
\ref{fig4}b and \ref{fig5}a move
into identical directions, their 
speeds (note the different time-steps 
$dt$) and orientations differ.)
Quantitatively, we thus obtained an 
experimental deflection angle of 
$\vartheta_\LL=12.1^\circ \pm 3.4^\circ$.

\begin{figure}[h]
\epsfxsize=\columnwidth
\epsfbox{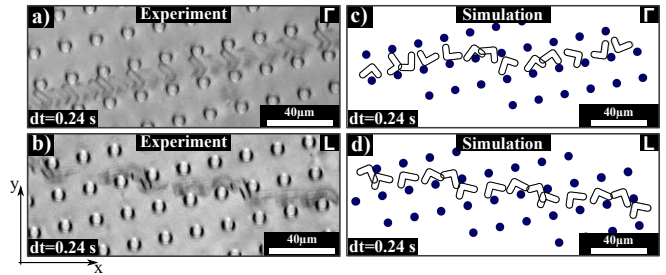}
\caption{
Left:
Same as in Fig. \ref{fig4}, but for {\boldmath$\alpha=12.5^\circ$}.
}
\label{fig6}
\end{figure}

Similarly, Fig. \ref{fig6} illustrates the experimentally 
and numerically observed particle motion for the
second of the above predicted ``most promising''
deflection angles, namely $\alpha=12.5^\circ$.
In contrast to $\alpha=15^\circ$, now the
agreement between experiment and theory is excellent
for the $\LL$-shaped particles, but somewhat
worse for the $\Gamma$-shaped particles.
Moreover, for the $\Gamma$-shaped particles 
there is again a second type of motion 
in the experiment without a theoretical 
counterpart, see Fig. \ref{fig5}b.
Alternatively, the motion of the $\Gamma$-shaped
particles predicted by the theory may be viewed as
a compromise between the two experimentally
observed types of motion.

We emphasize once again that the main purpose of our 
numerical simulations was to forecast a promising 
layout for our subsequent experimental realization 
of a microfluidic chiral selector device.
In view of Fig. \ref{fig3}, our theoretical approach was 
clearly very successful in this respect.
Accordingly, a further improvement of the
simulated model \cite{f1}
is beyond the scope of our present work.
Yet, since the quantitative agreement between simulations
and experiments in Figs. \ref{fig2}-\ref{fig6} is 
clearly not fully perfect,
it is worthwhile to point out some plausible reasons.
First of all, it is obvious, both from the simulations
and the experiments, that quite small changes
of $\alpha$ may lead to quite notable changes 
of the particle velocity with respect to 
direction, modulus, and the coexistence (or not) 
of more than one ``type of motion''.
Numerically, a comparably sensitive dependence
of the particle velocities was also observed
for most of the other model parameters.
One therefore expects that even quite small
imperfections 
of the theoretical model
may easily result in very significant 
errors of the predicted particle motion.
Indeed, our model \cite{f1} involves several 
approximations of this kind, most notably:
(i) Along the $z$-direction, the microfluidic
device from Fig. \ref{fig1} is bounded by a ``bottom-''
and a ``top-plane'', both of which are, however,
not perfectly planar but rather exhibit
notable imperfections/roughness/defects.
Every now and then, a traveling particle 
``bumps'' into one of them (due to thermal 
diffusion in $z$-direction etc.)
with the result of a quite notable, sudden
``deflection'' (even visible under the 
microscope).
(ii) The experimental post array deviates from a perfect 
square lattice and the single posts 
are not at all of perfect cylindrical 
shape.
(iii) While the force exerted by the
fluid flow on the particles is estimated in the
theoretical model by means of the fluid flow
in the absence of the particles, in reality
there will be quite notable corrections
due to hydrodynamic interactions between 
particles and microstructure.
In principle, it is not difficult to 
account for all those effects (i)-(iii) in
a theoretical model dynamics, but its numerical
simulation would require (among others)
to solve the full,
three-dimensional hydrodynamics at every 
time step, which goes far beyond of what 
is feasible in practice.
On the other hand, the qualitative and 
quantitative agreement between theory 
and experiment in Figs. \ref{fig2}-\ref{fig6} is 
still remarkably good in view of all those
approximations (i)-(iii).

{\em Conclusion:}
The main point of our paper was an experimental 
proof-of-principle that chiral particles 
can be separated without employing any kind of
chiral selector structure but rather by 
exploiting a dynamical chiral symmetry 
breaking.
In order to predict a suitable design of the 
experimental set-up, a simple theoretical
model dynamics has been introduced and 
numerically solved.
Any further improvement of the various
approximations of this model itself are 
beyond the scope of our present work.
As illustrated by Fig. \ref{fig3}, not a 
single particle in the experiment exhibited
a deflection angle with a ``wrong'' 
sign, i.e. the chiral separation fidelity
was extremely high.
A more systematic exploration and 
optimization of the proposed microfluidic 
device, e.g., with respect to
some specific technological applications or
a quantitative comparison with other
methodologies will certainly be an important issue
in the future, but goes beyond the objectives
of our present study.

Besides the chiral separation {\em per se},
one of our main findings was that the
motion of the chiral particles through
the periodic microstructure depends very
sensitively on the specific values of
many dynamical and structural system 
parameters.
On the one hand, the particle motion is thus
very difficult to predict qualitatively by
means of simple intuitive arguments and with
high quantitative accuracy by means of 
numerical simulations.
On the other hand, this sensitivity underlines
the practical efficiency and variability of 
the experimental concept.
The most promising future direction is now
to downscale the generic principles behind
our present approach into the 
realm of chiral nano-particles 
and  molecules.

\begin{center}
\vspace{-5mm}
---------------------------
\vspace{-4mm}
\end{center}
This work was supported by the German Research Foundation (DFG) within the Collaborative Research Center SFB 613.


\begin{thebibliography}{20}
\bibitem{f2} Note that for every $\alpha$-value
in Fig. \ref{fig1} one has to fabricate a new microstructure.
\end{thebibliography}
\end{document}